\documentclass[10pt,aps.prl,twocolumn,superscriptaddress,floatfix]{revtex4-1} 

\usepackage{amsmath}
\usepackage{graphics}
\usepackage{xcolor}
\usepackage{caption}
\usepackage{subcaption}
\usepackage{float}
\usepackage{array}
\usepackage[margin=1.0in]{geometry}
\usepackage{ulem}

\begin{document}
\title{Effect of Experimental Parameters on Optimal Reflection of Light from Opaque Media}
\author{Benjamin R. Anderson$^1$, Ray Gunawidjaja$^1$, and Hergen Eilers$^{*}$}
\affiliation{Applied Sciences Laboratory, Institute for Shock Physics, Washington State University,
Spokane, WA 99210-1495}
\date{\today}

\email{eilers@wsu.edu}

\begin{abstract}
Previously we considered the effect of experimental parameters on optimized transmission through opaque media using spatial light modulator (SLM) based wavefront shaping. In this study we consider the opposite geometry in which we optimize reflection from an opaque surface such that the backscattered light is focused into a spot on an imaging detector.  By systematically varying different experimental parameters (genetic algorithm iterations, bin size, the SLM active area, target area, spot size, and sample angle with respect to the optical axis) and optimizing the reflected light we determine how each parameter affects the intensity enhancement. We find that the effects of the experimental parameters on the enhancement are similar to those measured for a transmissive geometry, but with the exact functional forms changed due to the different geometry and the use of a genetic algorithm instead of an iterative algorithm. Additionally, we find preliminary evidence of greater enhancements than predicted by random matrix theory suggesting a possibly new physical mechanism to be investigated in future work.

\vspace{1em}


\end{abstract}

\maketitle

\vspace{1em}

\section{Introduction}
In 2007 Vellekoop and Mosk demonstrated the ability to focus light through an opaque medium using wavefront shaping via a liquid crystal on silicon spatial light modulator (LCOS-SLM) \cite{Vellekoop07.01}. This experimental observation was the first demonstration of Freund's 1990 prediction that wavefront shaping could be used to control the optical properties of opaque media \cite{Freund90.01}. After Vellekoop and Mosk's success with focusing light through a scattering system, the technique of wavefront shaping has been applied to numerous applications, including: control the spatio-temporal characteristics of random lasers \cite{Leonetti13.02,Leonetti12.02,Bachelard14.01,Bachelard12.01}, enhance fluorescence microscopy \cite{Vellekoop10.01,Wang12.01,Ghielmetti12.01}, achieve spectral control of a broadband light source \cite{Small12.01,Park12.02,Paudel13.01,Beijnum11.01}, compress ultrashort pulses \cite{McCabe11.01, Katz11.01}, control polarization \cite{Park12.01,Guan12.01}, achieve perfect focusing \cite{Vellekoop10.01, Putten11.01}, phase conjugation of fluorescence in turbid tissue \cite{Vellekoop12.01}, tunable beam splitters \cite{Huisman14.01}, spatial control of second-harmonic light \cite{Rodriguez12.01,Yao12.01}, control of single-photon Fock-state propagation \cite{Huisman14.02}, control of photocurrent in disordered photovoltaics \cite{Liew15.03}, focusing through dynamic tissue \cite{Wang15.01}, improving free-space optical communication \cite{Ren14.01}, image projection through disordered media \cite{Conkey12.03}, three dimensional microscopy \cite{Yang12.01}, improved optical coherence tomography \cite{Jang13.01}, creation of an ultrafast nanophotonic switch \cite{Strudley14.01}, optical control of excitation waves in cardiac tissue \cite{Burton15.01} and enhance astronomical/biological imaging \cite{Mosk12.01,Stockbridge12.01,Niv15.01} (with further improvements to biological imaging developed using the supplementary technique of photoacoustic wavefront shaping (PAWS) \cite{Lai14.02,Chaigne14.01,Tay14.02,Gigan14.01,Tay14.01,Lai14.01,Staley13.01}). In addition to these different applications, the technique of wavefront shaping to control the optical properties of opaque media has been proposed as a viable implementation method for optical physically unclonable functions \cite{Goorden14.01, Horstmeyer13.01,Skoric13.01,Anderson14.06,Eilers14.01,Anderson15.06}.

In general, physically unclonable functions (PUFs) are systems with a large number of randomly distributed degrees of freedom such that it is practically impossible to reproduce the system, thus making them unclonable \cite{Johnston01.01,Goorden14.01,Skoric13.01,Pappu02.01,Tuyls07.01}.  This random distribution can be interrogated using different techniques (electrical signals, acoustic waves, optical waves, etc.) to produce a unique signature which corresponds to the system's specific realization of disorder.  In an ideal PUF any change to the random distribution will result in the unique signature changing, thus giving evidence of a different distribution.  These properties -- irreproducibility and unique signatures -- make PUFs attractive in the fields of secure authentication and cryptography.

In the case where optical techniques are used to probe the system, the PUF is known as an optical physically unclonable function (O-PUF). The most common O-PUF is a scattering system, such as a nanoparticle (NP)-doped solid matrix. In these systems the position of all the scatterers are the degrees of freedom and any change to the NP positions will result in a different optical signature.  To date, the majority of research on O-PUFs has focused on passive determinations of the O-PUF's optical signature, with the primary method being to measure the speckle pattern of a laser beam scattered from/through the PUF \cite{Pappu02.01,Cowburn08.01,Seem09.01,Buchanan05.01,Shih14.01,Bromberg15.01}.  In this technique the speckle pattern (or a mathematical transformation of the pattern \cite{Pappu02.01}) becomes the unique optical signature.

While the passive method of measuring the PUF's speckle pattern is a strong method of authenticating a PUF's veracity, an additional layer of security can be introduced by utilizing wavefront shaping, which in essence turns the PUF into a ``lock'' and the shaped wavefront into a ``key''.  Only by probing the correct PUF with the correct key will a predetermined optical response result.  Typically this optical response is a focused spot on a detector with any other wavefront (or PUF) resulting in a speckle pattern on the detector \cite{Goorden14.01, Horstmeyer13.01,Skoric13.01,Eilers14.01}.

In order to successfully implement wavefront shaping as a probe of an O-PUF for secure authentication we must first understand the underlying factors which impact the system's performance.  To this end we previously considered how different experimental parameters effect optimal transmission through opaque media and developed a wave propagation model to describe the observed experimental dependancies \cite{Anderson14.06}. While optimized transmission is one method for implementing O-PUFs, an alternative method is to use optimized reflection. In optimized reflection both the optical probe and corresponding response share the same optical path, with the optical response reflected back from the O-PUF.  This geometry is advantageous, as it is not always possible to place a detector in a position to measure transmission through the O-PUF, but it is simple to measure the reflected light. 

Given that the reflective geometry is beneficial for implementing O-PUFs, we previously used it to measure the stability of the wavefront-sample coupling under sample translation and rotation \cite{Anderson15.06}. These measurements used image correlation coefficients to determine how sensitive the wavefront-sample coupling was to the sample being moved. However, the influence of different experimental parameters on the optimization efficiency for the reflective geometry has yet to be explored. Therefore, in this study we consider the influence of different experimental parameters on optimizing reflection from opaque media. These parameters include number of algorithm generations, bin size, active SLM area, target area, sample position along the optical axis, and sample angle. We find that while the optimization efficiency's dependence on these experimental parameters is similar to the dependencies found in our study on the transmissive geometry using an iterative optimization algorithm, there are unique differences due to the reflective geometry and use of a genetic algorithm. 

\section{Background}
Before discussing our current experiments in a reflective geometry, we first provide an overview of previous work on optimized transmission to provide a context for the various parameters, metrics, and models used. The first experimental study on wavefront shaping to optimize transmission of light through an opaque media was performed in 2007 by Vellekoop and Mosk in which they focused light transmitted through a TiO$_2$ sheet into a focal point \cite{Vellekoop07.01}.  In their study they defined the metric of optimization to be the intensity enhancement $\eta$, which is given by \cite{Vellekoop08.01,Vellekoop07.01}

\begin{align}
\eta \equiv \frac{I}{\langle I_0\rangle},
\end{align}
where $I$ is the intensity in the target area after optimization and $\langle I_0 \rangle$ is the average intensity in the target area before optimization. Using an optical analogue to electron conduction in a disordered wire based on random matrix theory (RMT)\cite{Beenakker97.01,Pendry90.01,Pendry92.01,Mello88.01,Mello88.02}, Vellekoop and Mosk predicted that the ideal intensity enhancement is given by 

\begin{align}
\eta=\frac{\pi}{4}(N-1)+1, \label{eqn:RMT}
\end{align}
where $N$ is the number of modulated wavefront segments.  Note that Equation \ref{eqn:RMT} only depends on the number of modulated wavefront segments and predicts that the enhancement should be independent of sample properties and system parameters \cite{Vellekoop08.02}.

After Vellekoop and Mosk's study, further research on optimized transmission was performed and a wide array of different enhancements was obtained for similar numbers of modulated wavefront segments \cite{Vellekoop07.01,Park12.01,Guan12.01,Popoff10.01,Cui11.01,Conkey12.01,Conkey12.02}. Initially these discrepancies were posited to be due to system noise and sample persistence time \cite{Vellekoop08.01,Yilmaz13.01}, where the persistence time $T_p$, is defined as the time during which a sample's speckle pattern remains unchanged --  typically on the order of hours for solid systems and miliseconds for liquids and living tissue \cite{Vellekoop07.01,Vellekoop08.01,Briers95.01}. 

 In order to correct for the effect of persistence time, Vellekoop and Mosk modeled an iterative optimization algorithm using RMT and found that the modified enhancement is given by \cite{Vellekoop08.01}:

\begin{align}
\eta \approx \frac{\pi}{4N}\left(\frac{1-e^{-NT_i/(2T_p)}}{e^{T_i/(2T_p)}-1}\right)^2, \label{eqn:RMTP}
\end{align}
where $T_i$ is the time for one algorithm iteration to complete. In the limit of infinite persistence time Equation \ref{eqn:RMT} and \ref{eqn:RMTP} only differ by an offset factor, while in the limit of small persistence times Equation \ref{eqn:RMTP} approaches zero. 

Using a similar modeling approach to Vellekoop and Mosk, Yilmaz \textit{et al.} modeled the effect of noise on optimization and predicted that the enhancement in the presence of noise behaves as

\begin{align}
\eta=\frac{\pi}{4}N\left(1-\frac{N}{R^2}\right), \label{eqn:RMTN}
\end{align}
where $R$ is the signal-to-noise ratio of the system.

Based on Equations \ref{eqn:RMTP} and \ref{eqn:RMTN} we would expect that the enhancement would only depend on the number of modulated segments, persistence time, iteration time, and the signal-to-noise ratio of the system. However, recently we  performed an extensive study of the effect of different experimental parameters on the enhancement and found that the bin size $b$, number of phase steps $M$, active SLM area $L^2$, target radius $r$, and spot size $d$ all affect the enhancement \cite{Anderson14.06}. While there is no closed form expression for the enhancement as a function of all five variables, the enhancement as a function of the number of bins is found to follow

\begin{align}
\eta=1+\eta_0\left[1-e^{-N/N_0}\right],
\end{align}
where $1+\eta_0$ is the asymptotic enhancement and $N_0$ is the 1/e number of bins.  Note that both $\eta_0$ and $N_0$ depend on the other four parameters.

Given the drastic difference between previous models of optimized transmission and the experimentally measured enhancement as a function of the different experimental parameters, we developed an alternative model to describe the observed effects which is based on wave propagation of a Gaussian beam with a random phase front. This model is called the random phase Gaussian beam model (RPGBM).  Initially, we modeled the effect of an opaque sample as adding a random phase front to a Gaussian beam, but neglected scattering effects to the beam's intensity profile at the exit surface of the sample, which lead to some discrepancies between the RPGBM and experiments \cite{Anderson14.06}. To correct for these shortcomings we later improved the RPGBM by including RMT modeling to describe the effect of the sample on the incident beams amplitude and phase \cite{Anderson15.06}.

While the RPGBM was initially formulated to model how different experimental parameters affect optimization of transmission through opaque media, it should be noted that there is a symmetry between transmission and reflection in which both the transmitted and reflected fields should behave similarly during wave propagation from the sample surface.  This symmetry implies that the predictions of the RPGBM should be independent of whether we perform optimized transmission or reflection. However, as we will see below, the choice of optimization algorithm and detector geometry does affect the parameter dependencies.

\section{Method}
To determine how different experimental parameters affect optimizing reflection we first prepare ZrO$_2$ NP doped polyepoxy nanocomposites to use as our scattering media. These nanocomposites are prepared as follows: first, spherical ZrO$_2$ NPs are synthesized by forced hydrolysis followed by calcination at 600 $^\circ$C for 1 hr \cite{Gunawidjaja13.01}. The NPs are then hydrophobized in a 1 vol\% solution of $n$-octadecyltriethoxysilane in toluene \cite{Gunawidjaja11.01}.  The hyrophobized ZrO$_2$ NPs are then dispersed in solution of bisphenol A diglucidyl ether (BADGE) and toluene, with a BADGE concentration of 20 mg/ml and a NP concentration of 10 wt\%.  Once mixed, the solution is sonicated to aid in dispersal.  Next, additional BADGE is added to yield the desired ZrO$_2$ concentration and the mixture is further sonicated to achieve a homogenous mixture and then the toluene is evaporated in vacuum. After mixing and evaporating we add an equivalent amount of diethylene triamine (DETA) curing agent and thoroughly mix the mixture. Once mixed the mixture is poured onto a 1'' $\times$ 1.5" glass slides and the nanocomposite coated glass slides are placed in an 80 $^\circ$C oven for 2 hours with the resulting films having thicknesses of 1 -- 3 mm and scattering lengths on the order of several microns.  Given that the sample thickness $L$, is much greater than the scattering length we conclude that our samples lie in the multiple scattering regime. 

\begin{figure}
 \centering
 \includegraphics{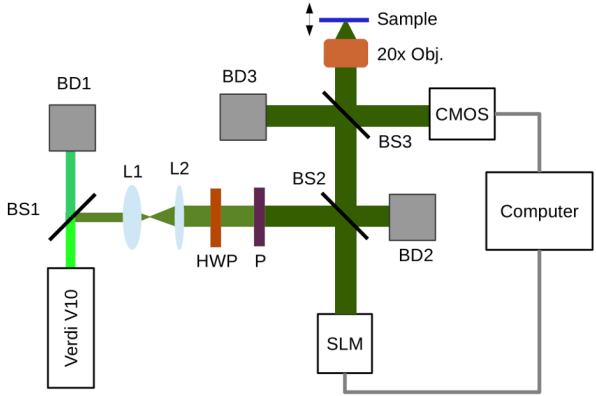}
 \caption{(Color Online) Schematic diagram of reflective geometry optimization setup. BD: Beam dump, L1: 40 mm lens, L2: 150 mm lens, BS1: 90:10 beamsplitter, BS2,BS3: 50:50 beamsplitter, HWP: half-waveplate, P: Polarizer.}
 \label{fig:setup}
\end{figure}

In order to optimize reflection from the samples we use an LCOS-SLM based wavefront optimization system in a reflection geometry, shown schematically in Figure \ref{fig:setup}.  The system consists of a Coherent Verdi V10 Nd:YVO$_4$ CW laser, a Meadowlark (formerly Boulder Nonlinear Systems) LCOS-SLM, a Mitutoyo 20$\times$ high working distance objective, a Thorlabs DCC1545 monochrome CMOS camera, and various focusing and polarization optics. To control sample alignment, we mount the sample on a custom sample holder connected to a Thorlabs translational stage and Newport rotation stage.  Optimization is controlled using a feedback loop between the SLM and CMOS detector with the optimization algorithm being a simple genetic algorithm \cite{Anderson15.02}.

\begin{figure}
\centering
\includegraphics{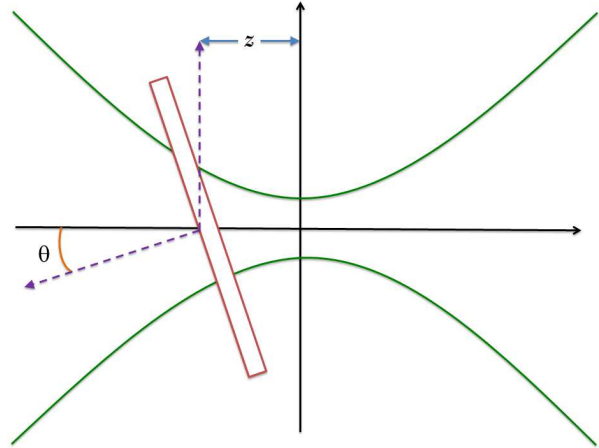}
\caption{(Color Online) Schematic of experimental geometry parameters with the Gaussian beam envelope overlaid. The angle $\theta$ is measured relative to the optical axis and the distance $z$ is measured relative to the focal point.}
\label{fig:geometry}
\end{figure}

When considering the effects of different experimental parameters on optimized reflection we chose the six main parameters:  number of algorithm generations $G$, bin size $b$, active SLM area $L^2$ \footnote{Note that the bin size and active SLM area determine the number of modulated bins on the SLM given by $N=(L/b)^2$.}, target area $A=\pi r^2$, where $r$ is the target integration radius, sample position along the optical axis $z$, and sample angle relative to the optical axis $\theta$. See Figure \ref{fig:geometry} for a schematic of the experimental geometry showing both $z$ and $\theta$. Using these six parameters we determine their effects on optimized reflection by performing optimization while varying one parameter at a time and holding all other parameters fixed.  In order to determine some of the covariant dependencies of the enhancement on the parameters, we perform the optimization experiments using several different parameter configurations  in which a second parameter acts as a sub-variable, e.g. measuring the enhancement as a function of sample angle for three different sample $z$ positions.

At this point we note that while in this paper we report results from ZrO$_2$ NP doped polyepoxy, the dependence of the enhancement on experimental parameters is applicable for any scattering sample. While not discussed here, our measured dependencies are found to be consistent for other scattering samples including: paper, polyurethane with dispersed NPs, Y$_2$O$_3$ ceramics, and ground glass. Additionally, while we use specific hardware (e.g. lenses, camera, SLM) for our experiments, the measured parameter dependencies should be independent of the hardware used, with the hardware only influencing the precise values of fit parameters.  This universality is due to our results being found to arise from wave propagation and scattering from disordered media, which are general phenomenon applicable in a wide range of experimental configurations.

\section{Results and Discussion}

\subsection{Number of Algorithm Generations}
While we previously used an iterative algorithm (IA) to optimize transmission through opaque media \cite{Anderson14.06}, in this study we use a simple genetic algorithm (SGA) \cite{Anderson15.02} as it is both faster and more resistant to experimental noise than the iterative algorithm.  One of the main differences between the IA and SGA is that the IA uses discrete phase steps, while the SGA is a stochastic algorithm that uses randomly generated phase values that are not discretized \cite{Anderson15.02}. This means that the number of phase steps $M$, considered previously in our transmission study, is no longer a valid parameter for testing.  Instead the SGA admits a new parameter, which is the number of algorithm generations $G$. 

\begin{figure}
\centering
\includegraphics{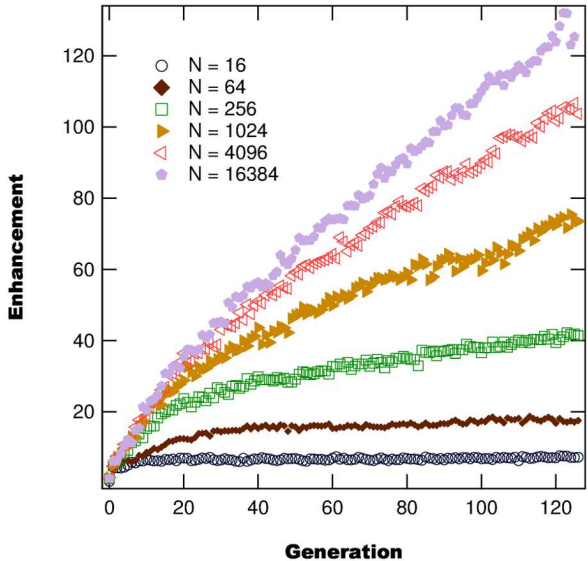}
\caption{(Color Online) Enhancement as a function of iteration for different bin numbers.}
\label{fig:gen}
\end{figure}

\begin{figure}
 \centering
 \includegraphics{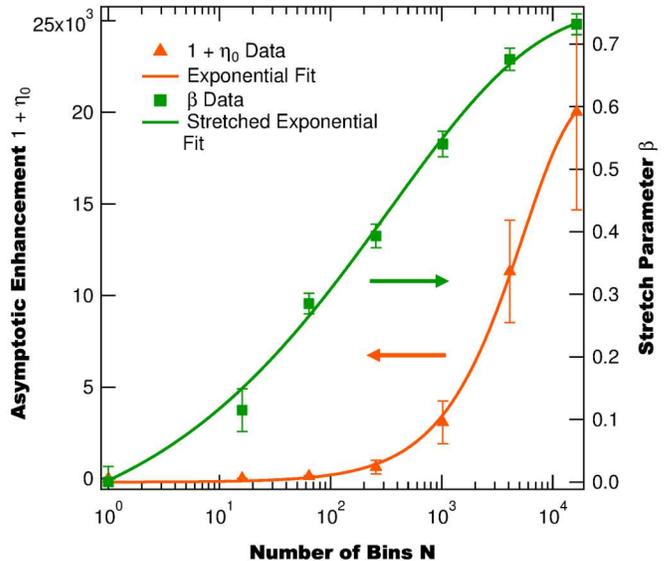}
 \caption{(Color Online) Asymptotic enhancement and stretch parameter as a function of different number of bins.  The asymptotic enhancement is found to behave as a simple exponential function and the stretch parameter behaves as a stretched exponential given by Equation \ref{eqn:str}.}
 \label{fig:etabin}
\end{figure}

To measure the effect of the number of algorithm generations on the enhancement we use the full SLM area ($L=512$),  place the sample surface at $z=0$ (giving a spot size of $d=1$ $\mu$m), use an integration radius of $r=2$, a sample angle of $\theta=0$, and six different bin sizes corresponding to total number of bins of $N=\{16,64,256,1024,4096,16384\}$. Figure \ref{fig:gen} shows the enhancement as a function of the number of generations for different total number of bins, with the enhancement found to follow a stretched exponential function given by,
\begin{align}
\eta(G)=1+\eta_0\left[1-e^{-(G/G_0)^\beta}\right], \label{eqn:gen}
\end{align}
where $1+\eta_0$ is the asymptotic enhancement, $G_0$ is the 1/e number of generations, and $\beta$ is the exponential stretch parameter.  Fitting the data in Figure \ref{fig:gen} to Equation \ref{eqn:gen} we determine the fit parameters for each total number of bins, which are tabulated in Table \ref{tab:gen} and displayed in Figure \ref{fig:etabin} as a function of the number of bins.

\begin{table}
\centering
\caption{Fit parameters from Equation \ref{eqn:gen} for different bin numbers.  While the asymptotic enhancement and stretch parameter are found to vary with the number of bins, the width parameter is found to be consistent for all bin numbers with an average value of $G_0 = 1.32(\pm 0.40)\times 10^5$.}
\label{tab:gen}
\begin{tabular}{|c|c|c|}
\hline
$N$  &  $1+\eta_0$  & $\beta$  \\ \hline
16384       &   $20000 \pm 10000$  &   $0.7315 \pm 0.0067$ \\
4096    &   $11300 \pm 5400$ &   $0.6760 \pm 0.0071$\\
1024   &   $3100 \pm 1200$ &   $0.5400 \pm 0.0082$\\
256   &   $630 \pm 170$ &   $0.3930 \pm 0.0074$\\
64    &   $137 \pm 26$ &   $0.2852 \pm 0.0066$\\
16 &   $16.9 \pm 1.1$   &  $0.115 \pm 0.014$\\ \hline
\end{tabular}
\end{table}

From Figure \ref{fig:etabin} and Table \ref{tab:gen} we find that the asymptotic enhancement as a function of the number of bins follows a simple exponential given by
\begin{align}
 1+\eta_0=\eta_M\left[1-e^{-N/N_{0,a}}\right], \label{eqn:assym}
\end{align}
where $\eta_M$ is the maximum enhancement attainable (e.g. enhancement as both $G$ and $N$ go to infinity) and $N_{0,a}$ is the 1/e number of bins for the asymptotic amplitude. Fitting the asymptotic enhancement in Figure \ref{fig:etabin} to Equation \ref{eqn:assym} we find that $\eta_M=21096 \pm 395$. Additionally, from Figure \ref{fig:etabin} and Table \ref{tab:gen}, we find that the exponential stretch parameter follows a stretched exponential given by,

\begin{align}
 \beta(N)=\beta_0\left[1-\exp\left\{-\left(\frac{N}{N_{0,b}}\right)^{1/3}\right\}\right],\label{eqn:str}
\end{align}
where $\beta_0$ is the asymptotic stretch parameter and $N_0$ is the 1/e number of bins. Fitting the stretch parameter in Figure \ref{fig:etabin} to Equation \ref{eqn:str} we determine that $\beta_0=0.7564\pm0.0081$ and $N_{0,b}=343\pm36$.

Based on these results we conclude that partitioning the SLM into smaller and smaller bins quickly reaches a point of diminishing returns, where further partitioning leads to only minor increases in the enhancement.  This can be seen as the asymptotic enhancement and stretch parameter are both near their maximal values for $N=16384$.  On the other hand we find that the number of generations remains very important out to large values of $G$. Namely, the 1/e number of generations from Figure \ref{fig:gen} is found to be $G_0 = 1.32(\pm 0.40)\times 10^5$ which is over 1000$\times$ larger than the maximum number of generations used in this study.

\begin{figure}
 \centering
 \includegraphics{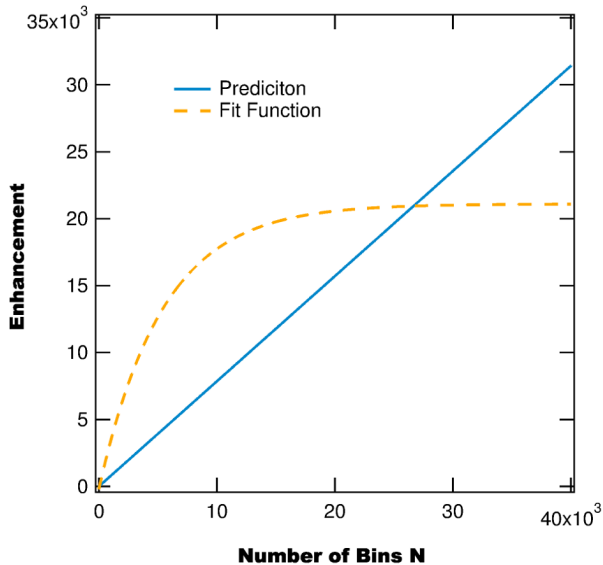}
 \caption{(Color Online) Asymptotic enhancement fit curve and enhancement prediction from Ref \cite{Vellekoop07.01}.}
 \label{fig:Apred}
\end{figure}

The last major finding when considering the effect of the number of iterations, is evidence of an apparent violation of the maximum enhancement predicted by random matrix theory \cite{Vellekoop07.01}. To demonstrate this violation we plot the asymptotic enhancement and the RMT predicted enhancement as a function of the number of bins in Figure \ref{fig:Apred}.  From Figure \ref{fig:Apred} we find that for bin numbers below approximately $N=28000$ the asymptotic enhancement is greater than the enhancement predicted by Equation \ref{eqn:RMT}, implying that the predicted RMT enhancement is incorrect.

While this finding may be evidence for new and unexplored physics, there are several possible alternative explanations. Firstly, we must note that we are comparing the RMT enhancement to the asymptotic enhancement determined by fits to the data in Figure \ref{fig:gen}, meaning that rather than measuring a direct enhancement violation, we are extrapolating from currently measured data.  Therefore there is a possibility that the enhancement at large numbers of generations could diverge from Equation \ref{eqn:gen}, which would make the current asymptotic enhancement invalid.  Ideally to test this possibility we would perform an optimization run with over 10$^5$ generations to see how the enhancement behaves.  However, this possibility is difficult to test as experimental noise and sample decoherence effects are found to become important after approximately 1000 generations, meaning that separating the effects of noise and decoherence from the underlying behavior is complicated.  To get around issues of noise and sample decoherence we are performing optimization modeling with the genetic algorithm for large numbers of generations to determine if the functional behavior remains the same.  Additionally we are working on accelerating the overall speed of the optimization system in order to allow more generations to be run before sample decoherence begins to be a major factor on the enhancement.

Another possible explanation to the enhancement violation is that the fundamental assumptions of the RMT enhancement prediction maybe violated in our system, thus making the comparison invalid. Therefore it is necessary to check the validity of these assumptions, which include: (1) the sample is disordered such that the reflection matrix coefficients $r_{mn}$, are statistically independent and obey a circular Gaussian distribution (2) all segments of the phase modulator contribute equally and (3) the number of segments $N$ is much less than the number of mesoscopic channels \cite{Vellekoop07.01, Vellekoopthesis,Vellekoop08.01, Beenakker97.01, Pendry92.01,Garcia89.01}.  Immediately we know that assumption 3 is not violated as the anomalous enhancement is seen at small $N$ rather than large $N$. Assumption 2 is most likely violated in our system as we use a TEM$_{00}$ Gaussian beam for our coherent source which has a spatially varying intensity.  This spatial variation will change the contribution of different SLM bins based on their position.  However, a violation of assumption 2 should produce a smaller enhancement, not a larger value.  Finally, the validity of assumption 1 is more difficult to ascertain as it requires measuring the reflection matrices of a large number of different configurations of disorder to determine the statistical properties of the matrices.  In order to determine whether or not assumption 1 is violated we are currently planning the necessary experiments to determine the reflection matrices of a wide array of different configurations of disorder \cite{Yu15.02}.

The final possible explanation for the anomalous enhancement  is that there are other factors that affect the enhancement which are not considered within the RMT model of optimization. These include noise \cite{Yilmaz13.01,Vellekoop08.01,Anderson14.06,Anderson15.02}, sample persistence time \cite{Vellekoop07.01,Vellekoop08.01,Yilmaz13.01,Anderson15.02}, and experimental configuration \cite{Anderson14.06}. While these effects are not considered by RMT, they are found -- in the case of optimized transmission -- to decrease the attainable enhancement and not improve it, which makes them appear to be poor candidates for the increased enhancement. However, there is a subtle difference between optimization of reflection and transmission, which may account for the increased enhancement.

In the case of the transmissive geometry the incident light couples into different transmission channels of the disordered system, such that the light exiting the sample is wholly determined by the transmission matrix. To optimize transmission to a specific mode requires shaping the incident wavefront such that the transmission matrix elements are matched to couple light into a target exit mode. In the ideal case of a perfectly shaped incident wavefront, the wavefront completely couples with the transmission matrix to give optimal transmission into the target mode.  However, in reality experimental limitations (SLM size, pixel size, SLM efficiency, objective numerical aperture, etc.) make the overlap of the experimental and optimal wavefront imperfect \cite{Vellekoop08.02,Goetschy13.01,Popoff14.01}.

For the reflective geometry, optimization is similar to the transmissive case, but with one key difference.  While the transmitted light from a disordered media is fully determined by the transmission matrix, in reflection there is a small amount of light which is ballistically reflected from the surface of the sample due to fresnel reflection ($\approx 5\%$) and therefore does not depend on the reflection matrix of the disordered system. As this light is not confined by the reflection eigenchannels of the disordered medium, it is possible to couple a large portion of the ballistically reflected light into the target area giving an enhancement greater than predicted by RMT (which ignores the ballistically reflected light). To better understand this effect and determine if it indeed explains the observed enhancement violation we are performing modeling using the RPGBM model with the inclusion of a small ballistically reflected wavefront in addition to the diffuse reflected light from the disordered sample.

\subsection{Bin Size}
\begin{figure}
\centering
\includegraphics{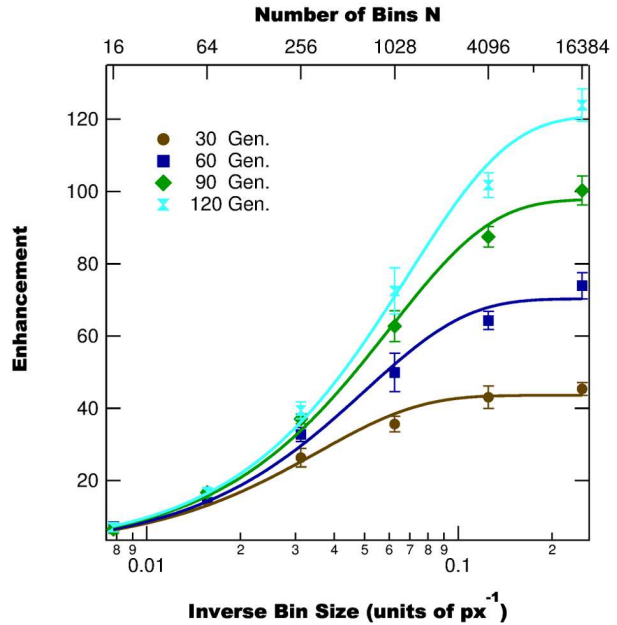}
\caption{(Color Online) Enhancement as a function of number of bins, which is found to be fit by a double exponential function.}
\label{fig:bin}
\end{figure}

Using the results of the enhancement as a function of generation for different numbers of bins we determine the effect of bin size on the enhancement. Figure \ref{fig:bin} shows the enhancement as a function of the inverse bin size for $G=\{30,60,90,120\}$ algorithm generations with the measured enhancement found to follow a stretched exponential function 

\begin{equation}
 \eta=1+\eta_0\left[1-\exp\left\{-\left(\frac{b_0}{b}\right)^{4/3}\right\}\right],\label{eqn:bins}
\end{equation}
where $1+\eta_0$ is the asymptotic enhancement and $b_0$ is a width parameter.

Fitting the data in Figure \ref{fig:bin} to Equation \ref{eqn:bins}, we determine the asymptotic enhancement and the bin size parameter for each different number of generations with the results tabulated in Table \ref{tab:bin}.   From table \ref{tab:bin} we find that as the number of generations increases the asymptotic enhancement increases and the width parameter decreases. The decrease in the width parameter with increasing number of generations corresponds to more bins being required to achieve the asymptotic enhancement. This means that more bins can be utilized before the enhancement saturates, which leads to greater possible enhancements.These results are consistent with Figure \ref{fig:gen} where the enhancement is shown to increase with number of generations and the effects of decreasing bin size are more pronounced for larger numbers of generations.

\begin{table}
\centering
\caption{Fit parameters from Equation \ref{eqn:bins} for the intensity enhancement as a function of bin size.}
\label{tab:bin}
\begin{tabular}{|c|c|c|}
\hline
$G$  &  $1+\eta_0$  & $b_0$   \\ \hline
30       &   $42.5 \pm 1.4$  &  $9.71 \pm 0.90$  \\
60    &   $67.5 \pm 2.3$  &   $7.84  \pm 0.91$ \\
90  &   $94.5 \pm 2.8$   &  $6.83 \pm 0.53$ \\ 
120 &  $117.6 \pm 3.6$ &  $6.08 \pm 0.57$  \\ \hline
\end{tabular}
\end{table}

These results are consistent with previous measurements of optimized transmission using a genetic algorithm -- where the enhancement behaved as a stretched exponential as a function of bin numbers \cite{Anderson15.02} -- suggesting that the reflective geometry does not influence how the bin size affects the optimization. While there is consistency between optimized reflection and transmission using a genetic algorithm, the results differ from optimized transmission using an IA, which behaves as a single exponential \cite{Anderson14.06}. This implies that the difference in the optimization's dependence on bin size is due to the algorithm and not the geometry.

\subsection{SLM Cropping}
\begin{figure}
\centering
\includegraphics{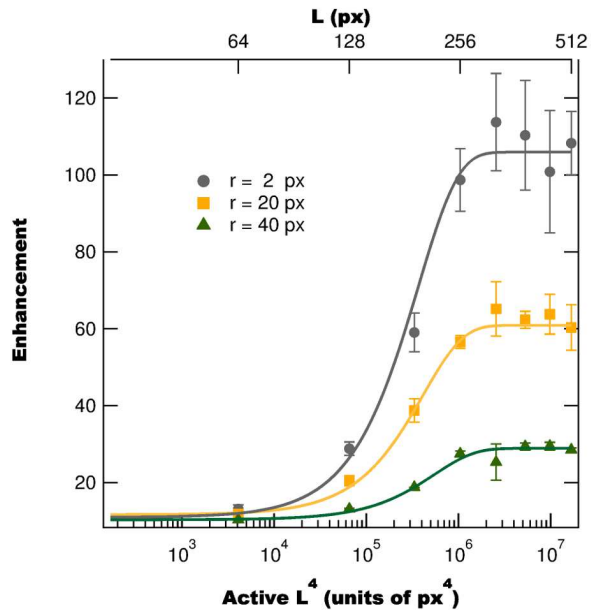}
\caption{(Color Online) Enhancement as a function of active side length.  The enhancement is found to scale exponentially with $L^4$.}
\label{fig:crop}
\end{figure}

Along with the bin size determining the total number of bins, the active side length $L$, of the SLM also changes the number of active bins.  To measure the enhancement's dependence on the active side length we use a spot size of $d=1$ $\mu$m, a bin size of 8 px, an angle of $\theta=0$, 120 generations, and three different target radii $r=$ \{2 px, 20 px, 40 px\}. Figure \ref{fig:crop} displays the enhancement as a function of active side length for different target radii, with the enhancement found to follow a super-Gaussian of the form,

\begin{equation}
\eta=1+\eta_0\left[1-\exp\left\{-\left(\frac{L}{\Delta L}\right)^4\right\}\right],\label{eqn:L}
\end{equation}
where $1+\eta_0$ is the asymptotic enhancement and $\Delta L$ is a width parameter. We note that Equation \ref{eqn:L} is functionally identical to the form observed for the transmission geometry, which is to be expected due to the symmetry of wave propagation between reflection and transmission. Fitting the data in Figure \ref{fig:crop} to Equation \ref{eqn:L} we determine the fit parameters for the three different target radii, with the results shown in Table \ref{tab:crop}.  From Table \ref{tab:crop} we find that as the target radius decreases the asymptotic enhancement increases and the Gaussian width is constant within experimental uncertainty. While the enhancement increase with decreasing target radius is expected, the observation of the width parameter remaining constant with target radius is unexpected and different to the behavior observed in the transmission geometry \cite{Anderson14.06}.

\begin{table}
\centering
\caption{Fit parameters from Equation \ref{eqn:L} for the intensity enhancement as a function of active area.}
\label{tab:crop}
\begin{tabular}{|c|c|c|}
\hline
$r$  &  $1+\eta_0$  & $\Delta L$  \\ \hline
2       &   $105.9 \pm 3.6$  &   $24.63 \pm 0.54$ \\
20    &   $60.9 \pm 2.5$ &   $25.38 \pm 0.53$\\
40 &   $28.9 \pm 2.6$   &  $27.2 \pm 2.2$\\ \hline
\end{tabular}
\end{table}

In the case of optimizing transmission through opaque media, a strong inverse relationship between the target radius and the width in Equation \ref{eqn:L} was observed. The inverse relationship between the Gaussian width and spot size is due to the Fourier relationship between the sample and detector planes \cite{Anderson14.06}. Upon initial consideration it seems intuitive that the Fourier relationship between the sample and detector planes should remain consistent between the two geometries; however, we observe that this is not the case. 

To understand this discrepancy we need to consider the different operations of the IA and SGA. The IA operates by sequentially optimizing each pixel. This typically manifests in an optimization curve in which the outer pixels have little influence on the optimization, while the central pixels produce the largest changes in the enhancement.  With regards to cropping, this method leads to the observation of the Fourier relationship between the cropping width and target radius as each pixel is optimized individually and only those within a certain width contribute to optimization.  However, in the case of the SGA all pixels are optimized simultaneously such that the Fourier relationship is not observed.  This simultaneous optimization leads to the target area not influencing the cropping width. Note that in case that the number of generations approaches $G_0$ we would anticipate that the phase mask will start approaching the same as found via the IA, resulting in the Fourier relationship being recovered. To test this prediction we are performing modeling in which the number of generations is set to very large values.

\subsection{Target Area}
\begin{figure}
\centering
\includegraphics{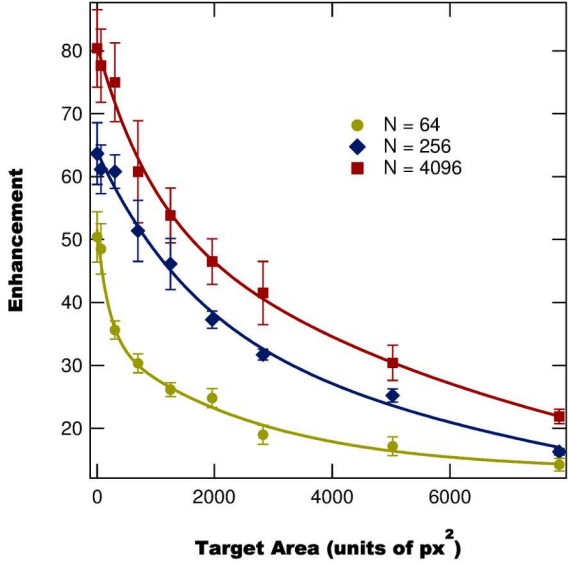}
\caption{(Color Online) Enhancement as a function of target area for different bin sizes and a spot size of 100 $\mu$m.}
\label{fig:areabin}
\end{figure}

\begin{figure}
\centering
\includegraphics{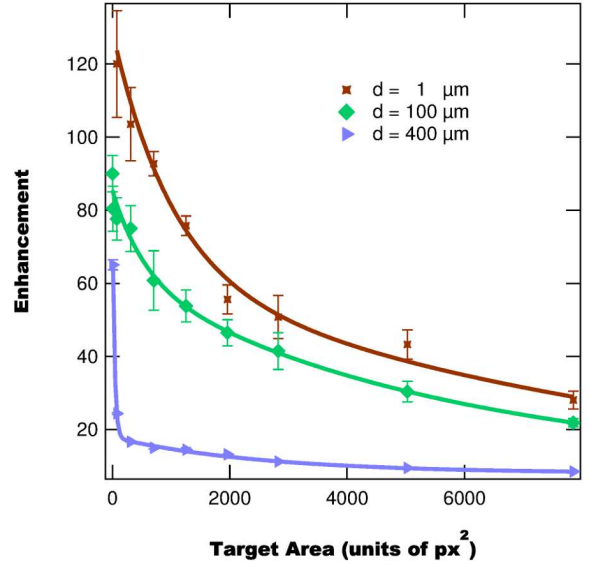}
\caption{(Color Online) Enhancement as a function of target area for different beam spot sizes and a bin size of 8 px.}
\label{fig:areaspot}
\end{figure}

Thus far we have considered optimization parameters related to the SLM (bin size and cropping) and overall operation of the SGA (number of generations).  In addition to these parameters, the SGA also depends on the target area being optimized on the detector.  We therefore consider the effect of changing the target area on the enhancement.  For these tests we use 120 generations, the full SLM ($L=512$), an angle of $\theta=0$, three different bin sizes $b$ = \{8 px, 16 px, 32 px\}, and three different spot sizes $d$ = \{1 $\mu$m, 100 $\mu$m, 400 $\mu$m\}.  Figure \ref{fig:areabin} shows the enhancement as a function of target area for different bin sizes and Figure \ref{fig:areaspot} shows the enhancement for different spot sizes.  From both figures we find that the enhancement as a function of target area behaves as the sum of two exponentials given by 

\begin{align}
\eta=1+\eta_1e^{-A/A_1}+\eta_2e^{-A/A_2}, \label{eqn:r}
\end{align}
where $\eta_1$ and $\eta_2$ are the exponential amplitudes for the slow and fast components, respectively, and $A_1$ and $A_2$ are the 1/e areas for the slow and fast components, respectively. This functional form is identical to that observed for the transmissive geometry. Using Equation \ref{eqn:r} we fit the data in Figures \ref{fig:areabin} and \ref{fig:areaspot} and tabulate the fit parameters in Table \ref{tab:area}.

\begin{table*}
\centering
\caption{Fit parameters from Equation \ref{eqn:r} for the intensity enhancement as a function of integration area for differnet bin sizes and spot sizes.}
\label{tab:area}
\begin{tabular}{|c|c|c|c|c|c|}
\hline
$d$  ($\mu$m)  & $N$ &  $\eta_1$  & $A_1$ (px$^2$)  &  $\eta_2$  & $A_2$ (px$^2$)  \\ \hline
100  & 64   &   $21.3 \pm 3.0$   &   $2607.7 \pm 1.4$   &  $18.3 \pm 5.0$  &  $198  \pm  1.5$   \\ 
100  & 256   &   $46.4 \pm 1.5$   &   $8760  \pm 30$  &  $22.11 \pm 1.45$  &  $970  \pm  64$\\
100  & 4096     &   $56 \pm 10$   & $7638.8 \pm 4.4$    & $25.9 \pm  5.4$  &  $925.3  \pm  2.1$ \\ \hline
 1     & 4096     &  $64.69 \pm 0.58$   &  $11110 \pm 140$    & $69.8 \pm 6.6$  &  $1021 \pm 13$ \\
100  & 4096     &   $56 \pm 10$   & $7638.8 \pm 4.4$    & $25.9 \pm  5.4$  &  $925.3  \pm  2.1$ \\
400  & 4096     &  $9.81 \pm 0.58$   &   $2570.1 \pm 1.5$   &  $68.35 \pm 0.95$  &  $34.0  \pm 2.8$   \\ \hline
\end{tabular}
\end{table*}

From Table \ref{tab:area} we find that as the number of bins increases the exponential amplitudes increase and the 1/e areas behave non-monotonically by first increasing when going from $N=64$ to $N=256$ and then decreasing when going from $N=256$ to $N=4096$.  The behavior of the amplitudes is expected as the enhancement increases with the number of bins.  However, the behavior of the 1/e areas is unexpected given their behavior perviously seen when in the transmissive geometry.  In the case of optimized transmission, the 1/e areas were found to decrease with increasing number of bins, eventually reaching a steady state value \cite{Anderson14.06}.  

While more modeling and experiments are required to precisely determine the difference in the 1/e area behavior with bin number between the transmission study and the current study, we note that preliminary experimental results using the iterative algorithm suggest that the increasing 1/e area with number of bins is related to the use of a genetic algorithm. To understand why the choice of algorithm should influence the 1/e area we consider the underlying physics of optimization as a function of spot size.

When considering optimization of a speckle pattern it is known that the average speckle size roughly correlates with a single reflection (transmission) mode \cite{Vellekoopthesis,Vellekoop07.01,Mosk14.01}. Therefore when optimizing an area on the order of a single speckle the optimization algorithm works to couple light into a single exit mode. As the target area increases the number of modes being optimized increases. This increase in the number of modes being optimized results in the influence of noise and sample decoherence being greater, leading to a decrease in the possible enhancement for larger areas.

Given the nature of the influence of sample decoherence on optimization (see Equation \ref{eqn:RMTP}), we anticipate that its influence will be more pronounced as the total optimization time increases. For the iterative algorithm the total optimization time scales with $N$, while for the Genetic algorithm the total optimization time is independent of $N$. Therefore it makes sense that by increasing $N$ the 1/e areas should decrease for the iterative algorithm as the effect of decoherence and noise will be greater at larger areas and longer times. However, for the Genetic algorithm the increase in $N$ does not influence the optimization time, so the influence of decoherence and noise is unchanged.  Additionally, the increase of $N$ allows for greater coupling into multiple modes, which results in the 1/e area increasing.

Along with measuring the effect of bin size on the enhancement as a function of target area, we also consider the effect of spot size on the enhancement's functional behavior.  From Table \ref{tab:area} we find that there is an inverse relationship between the spot size and the 1/e areas, which can also be seen in Figure \ref{fig:areaspot} where the enhancement curves narrow as the spot size increases. This effect arises due to the Fourier relationship between the sample and detector planes.

\subsection{Sample Position}
\begin{figure}
\centering
\includegraphics{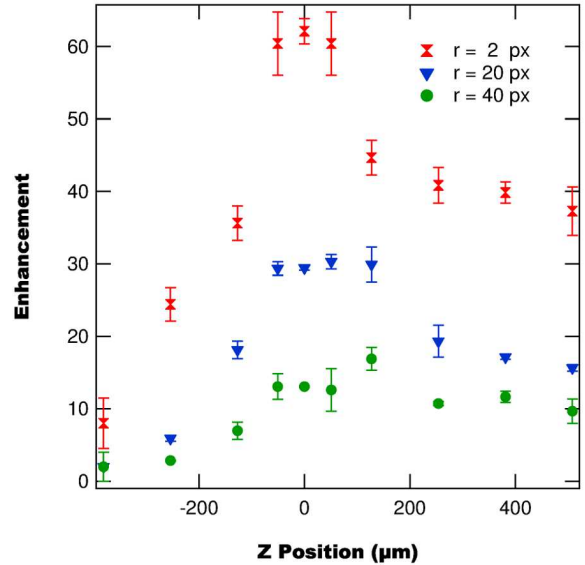}
\caption{(Color Online) Enhancement as a function of translation along the optical axis }
\label{fig:zpos}
\end{figure}

Along with the SLM and detector parameters, the enhancement is also influenced by changing the beam-sample intersection.  This is achieved by either moving the sample along the optical axis or rotating the sample such that the reflection is no longer normal to the surface of the sample.  We begin studying the effect of the beam-sample intersection on the enhancement by considering the sample's position along the optical axis with the objective's focal point defined as $z=0$.  For these tests we use the full SLM area, an angle of $\theta=0$, a bin size of 8 px, 60 generations, and three target radii of $r$ = \{2 px, 20 px, and 40 px\}. Figure \ref{fig:zpos} shows the enhancement as a function of position along the optical axis with the beam focal point defined as $z=0$ and positive $z$ positions corresponding the the sample being closer to the lens than the focal point.  From Figure \ref{fig:zpos} we find that the enhancement is not symmetric about the focal point and that it peaks near the focal point. This result differs from the transmissive case, in which the enhancement was found to be symmetric about the focal point and had its peak at positions away from the focal point.  

\begin{figure}
 \centering
 \includegraphics{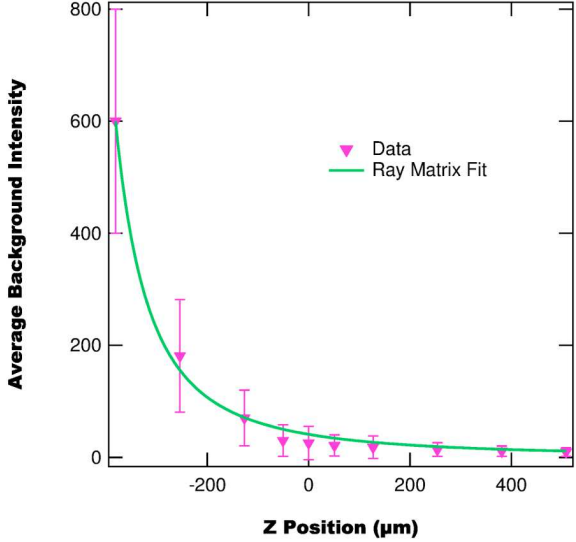}
 \caption{(Color Online) Average background intensity as a function of $z$ position with a fit to Equation \ref{eqn:back}.}
 \label{fig:backZ}
\end{figure}

The mechanism of this difference in behavior between the two geometries is related to the positioning of the target objective.  In the transmissive geometry the target objective remains at a fixed distance of one focal length from the sample such that no matter the beam size incident on the sample the transmitted light is collimated onto the detector.  However, in the reflective geometry the focusing objective also acts as the target objective such that the distance between the target objective and sample changes when moving the sample.  This leads to the back scattered light having some divergence/convergence from the objective for sample positions other than $z=0$, implying that the unoptimized intensity in the target area depends on the sample position.  To demonstrate this we plot the average unoptimized intensity in the target area as a function of position in Figure \ref{fig:backZ}.

To quantify the effect of the distance between the sample and objective we use ray matrices \cite{Saleh91.01,Menzel07.01} to determine the beam size at the detector plane. Using a simple model consisting of the reflected beam traveling a distance $x$ from the sample to an objective with focal length $f$, followed by propagation over a distance $L$ to the detector, we can determine the beam size at the detector to be

\begin{align}
 w_d=\left[1-L\left(\frac{1}{x}-\frac{1}{f}\right)\right]w_0,
\end{align}
where $w_0$ is the initial beam width focused onto the sample. Assuming that the power $P_0$, is unchanged with sample position we can write the background intensity on detector as
\begin{align}
 \langle I_0 \rangle &=\frac{P_0}{\pi w_d^2} \nonumber
 \\ &=\frac{P_0}{\pi\left[1-L\left(\frac{1}{x}-\frac{1}{f}\right)\right]^2w_0^2 }. \label{eqn:back}
\end{align}
Using Equation \ref{eqn:back}, we fit the average unoptimized intensity in Figure \ref{fig:backZ} and find them to be in good agreement.  

Given that the average unoptimized intensity increases drastically after the sample passes the objectives focal point, it is obvious that the enhancement, which is inversely proportional to $ \langle I_0 \rangle $, should decrease as the sample is translated. 


\subsection{Sample Angle}
\begin{figure}
\centering
\includegraphics{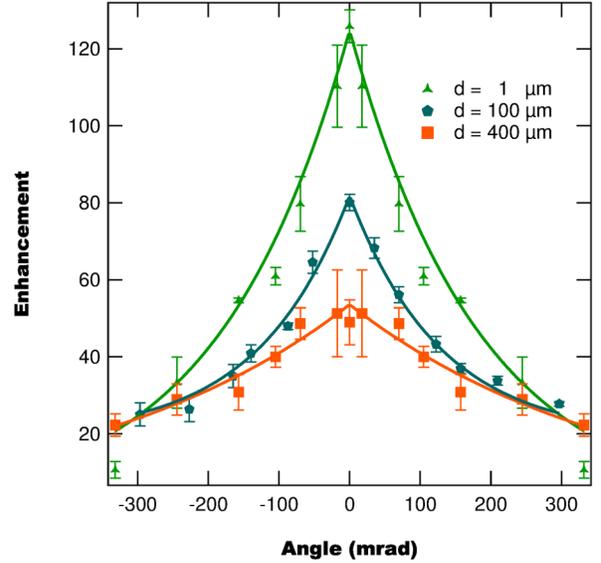}
\caption{(Color Online) Enhancement as a function of sample angle for three different spot sizes. The enhancement is found to be invariant under translation for an angle of approximately 250 mrad (14.3$^\circ$).}
\label{fig:angle}
\end{figure}

Finally, we consider the effect of the sample angle on the enhancement.  For these measurements we use a bin size of 8 px, 120 generations, a target radius of $r=2$, the full SLM area, and three different spot sizes $d$ = \{1 $\mu$m, 100 $\mu$m, and 400 $\mu$m\}.  Figure \ref{fig:angle} shows the enhancement as a function of sample angle for three different spot sizes, with the enhancement found to follow a simple exponential function given by

\begin{align}
\eta=\eta_0e^{-|\theta|/\theta_0}, \label{eqn:angle}
\end{align}
where $\eta_0$ is the zero-degree enhancement and $\theta_0$ is the 1/e angle.  From Figure \ref{fig:angle} we find that all three enhancement curves intersect at approximately 250 mrad (14.3$^\circ$) and that the 1/e angle, tabulated in Table \ref{tab:angle}, increases with spot size. Both these results are unexpected, as our previous study on the effect of sample rotation on the wavefront-sample coupling found that the wavefront-sample interaction is more stable for smaller spot sizes \cite{Anderson15.06}. These previous results would suggest that the 1/e angle should decrease with increasing spot size, which would in turn lead to the different enhancement curves not intersecting, implying that there should be no iso-enhancement angle.

\begin{table}
\centering
\caption{Fit parameters from Equation \ref{eqn:angle} for the intensity enhancement as a function of angle.}
\label{tab:angle}
\begin{tabular}{|c|c|c|}
\hline
$d$($\mu$m)  &  $\mathbf{\eta_0}$ & $\theta_0$(mrad)  \\ \hline
  1  &   $129.7 \pm 9.8$  &   $184.5 \pm 5.7$ \\
100  &   $69.3 \pm 1.1$    &   $296 \pm 9.1$\\
400  &   $53    \pm  23$    &  $370 \pm 46$\\ \hline
\end{tabular}
\end{table}

In order to better understand and explain these effects we are currently in the process of performing modeling using the modified RPGBM \cite{Anderson15.06}, with the spot size and sample angle influence on the wavefront taken into account. While further research is underway to fully explain the influence of the sample angle on enhancement, one explanation of the observed dependence is related to the nature of a focusing Gaussian beam and scattering from an angular surface. 

When light backscatters from a scattering media the light is not reflected in a collimated beam, but is instead scattered with some angular distribution $f(\phi;\theta)$, where $\phi$ is the outgoing angle measured relative to the optical axis and $\theta$ is once again the angle of the sample shown in Figure \ref{fig:geometry}. Not only does this distribution depend on the incoming and outgoing angles, it also depends on the wavefront of the incident light.  For a well collimated beam the wavefront is relatively flat and all rays are parallel. This means that when the beam scatters from the surface all rays will follow roughly the same distribution, leading to a tighter reflected pattern.  However, when the incident wavefront is not flat, with the different rays either diverging or converging, the resulting reflected pattern will have a wider angular distribution. This observation is simplified when considering reflection from a mirror: a collimated beam reflecting from a mirror remains collimated, while a converging or diverging beam incident on the mirror will result in the far-field beam diverging. 

The end result of the wider angular distribution is to allow more of the reflected light to be collected as the angle $\theta$ increases. This result is once again simplified by considering a mirror in place of the sample.  As the mirror is rotated the reflected beam will eventually be rotated to the point where it no longer reflects into the objective.  However, for a larger angular distribution, the angle at which the beam misses the focusing objective is greater, giving rise to a broader angular distribution as measured by the detector.  

These angular effects come into play when looking at the different beam spot sizes, as a Gaussian beam near its waist has a flat wavefront, while further from its waist, where the spot size is larger, the wavefront is converging. This difference in the incident wavefront as a function of spot size leads to the observed behavior in the enhancement as a function of sample angle.

\subsection{Comparing Reflection and Transmission}
We have thus far considered the influence of experimental parameters on optimized reflection using a simple genetic optimization algorithm.  At this point we summarize our results and make comparisons to the results for the transmissive geometry \cite{Anderson14.06}. Table \ref{tab:diff} lists the enhancement's dependencies on the different experimental parameters for the different geometries as well as the proposed mechanism leading to different behavior between the two geometries.

\begin{table*}[t]
\centering
 \caption{Enhancement dependence on experimental parameters for transmission and reflection geoemtries with proposed mechanisms for differences.}
 \label{tab:diff}
 \begin{tabular}{|>{\centering\arraybackslash}m{2.2cm}|>{\centering\arraybackslash}m{4cm}|>{\centering\arraybackslash}m{4cm}|>{\centering\arraybackslash}m{4cm}|}
 \hline
   \textbf{Experimental Parameter} & \textbf{Transmission}\cite{Anderson14.06}  & \textbf{Reflection}  & \textbf{Mechanism of Difference} \\   \hline 
  Algorithm Generations $G$  & -- & $1+\eta_0\left[1-e^{-(G/G_0)^\beta}\right]$ & n/a \\ 
   & & & \\ 
  Phase Steps $M$ & $1+\eta_0\cos^p\left(\frac{\pi}{2M}\right)$ & -- & n/a \\ 
     & & & \\ 
  Bin size $b$ & $1+\eta_0\exp\left\{-\left(\frac{b_0}{b}\right)^2\right\}$ & $1+\eta_0\left[1-\exp\left\{-\left(\frac{b_0}{b}\right)^{4/3}\right\}\right]$ & Fundamental difference in algorithms \cite{Anderson15.02} \\ 
     & & & \\ 
  Active SLM Side Length $L$ & $1+\eta_0\left[1-\exp\left\{\left(\frac{L}{\Delta L}\right)^4\right\}\right]$ ($\Delta L$ decreases with increasing $A$) & $1+\eta_0\left[1-\exp\left\{\left(\frac{L}{\Delta L}\right)^4\right\}\right]$ ($\Delta L$ is invariant with $A$) & Simultaneous vs Sequential Optimization\\ 
     & & & \\ 
  Target Area $A$& $1+\eta_1e^{-A/A_1}+\eta_2e^{-A/A_2}$ ($A_1,A_2$ decreases with $N$) & $1+\eta_1e^{-A/A_1}+\eta_2e^{-A/A_2}$ ($A_1,A_2$ increases with $N$) & Noise and decoherence resistance of algorithm\\ 
     & & & \\ 
  Sample Position $z$& Symmetric two-peak function with centers at $z\neq0$  & Asymmetric peak function with center at $z=0$  & Difference in detector geometry  \\ 
     & & & \\ 
  Sample Angle $\theta$& -- & $\eta_0e^{-|\theta|/\theta_0}$ & n/a \\ 
     & & & \\
  \hline
 \end{tabular}
\end{table*}

From Table \ref{tab:diff} we find that the functional form of the bin size dependence of the enhancement changes between the transmission based study and the current reflection based study. This difference is due to using a different algorithm, as it has been previously observed that a genetic algorithm produces a different bin size dependance than the iterative algorithm \cite{Anderson15.02}. Additionally, the functional dependence on the bin size for both the transmissive and reflective geometry is different than predicted by RMT, with the underlying mechanism related to beam propagation effects and the influence of experimental noise \cite{Anderson14.06}.

While we find that the functional form of the bin size dependence changes due to the different algorithms, we find that the functional form of the enhancement on the active SLM area and target area are the same between geometries. This observation is expected as these functional dependancies are related to beam propagation effects, which are independent of the geometry. However, despite being functionally the same we find that the fit parameters behave differently depending on the choice of algorithm, with the difference in target area dependence due to the algorithms' differing resistances to noise and sample decoherence \cite{Anderson15.02} and the difference in the cropping parameter dependence being due to the simultaneous optimization of the SGA.  These results show that the genetic algorithm is able to more successfully optimize large target areas than the iterative algorithm, with the fundamental difference being the genetic algorithms ability to resist the larger noise and sample decoherence as well as the simultaneous optimization versus pixel-by-pixel optimization.

Lastly from Table \ref{tab:diff} we find that the enhancement as a function of sample position along the optical axis is drastically different depending on the geometry.  For the transmissive case the enhancement is found to follow a symmetric two-peak function with the peak centers located at a non-zero $z$ position. However, for the reflective geometry the enhancement is found to follow an asymmetric single peak function with the peak occurring at $z=0$.  This difference arises due to the different detector geometries.  In the transmission case the the collection objective is placed at a fixed distance from the sample exit surface, such that the changes in the enhancement arose due to the changing spot size and not due to changes in the collection optics.  However, in the reflective case the focusing and collection objective are the same, such that a change in the sample positioning results in a change of the distance between the collection objective and the sample surface.  This change in positioning changes the beam propagation characteristics and leads to the different behavior of the enhancement observed for the reflective geometry.

\section{Conclusions}
Spatial light modulator controlled reflection from opaque media is a strong candidate for implementing optical physically unclonable functions for use in secure authentication. To help facilitate these applications it is necessary to understand the effects of different experimental parameters on an optimization algorithm's ability to enhance reflection.  In this study we measure the effects of six different experimental parameters on optimization, with the parameters being the number of algorithm generations, bin size, active SLM area, target area, sample position along the optical axis, and sample angle relative to the optical axis.

From these reflection based measurements we find that the enhancement as a function of different parameters behaves similarly to our previous study on the transmission geometry using an IA, with the majority of parameter effects based in the Fourier relationship between the sample and detector planes.  However, despite a large number of similarities, we find that for some parameters the reflective geometry and SGA produce different functionalities than previously observed for the transmission geometry and IA. These functional differences are primarily due to the different operations of the SGA and IA, as well as the positioning of the target objective. In order to better understand these differences and provide a predictive model of the parameter dependencies we are currently developing the RPGBM to account for the target objectives positioning and implementing the genetic algorithm into the optimization scheme.

Additionally from these measurements we find evidence of possible enhancements larger than predicted by RMT. This result suggests the possibility of new physical mechanisms currently ignored by RMT for scattering from disordered media.  We propose that the most likely candidate for this mechanism is Fresnel reflection from the air-sample interface, which is ballistically reflected and currently ignored by RMT.

\acknowledgments
This work was supported by the Defense Threat Reduction Agency, Award No. HDTRA1-13-1-0050, through Washington State University.

\newpage

\bibliographystyle{osajnl}
\bibliography{PrimaryDatabase,ASLbib}

\end{document}